\documentclass[twoside,twocolumn,aps,prb,preprintnumbers,floatfix,showpacs,superscriptaddress,citeautoscript,]{revtex4-1}

\usepackage{amsmath}
\usepackage{amssymb}
\usepackage{color}
\usepackage{graphicx}
\usepackage{tabularx}
\usepackage{subfigure}
\usepackage{dcolumn}
\usepackage{times}

\newcommand{\eV}[0]{\text{eV}}

\newcommand{\BT}{BaTiO$_3$}
\newcommand{\ST}{SrTiO$_3$}
\newcommand{\PT}{PbTiO$_3$}

\newcommand{\sect}[1]{Sect.~\ref{#1}}
\newcommand{\fig}[1]{Fig.~\ref{#1}}
\newcommand{\Fig}[1]{Figure~\ref{#1}}
\newcommand{\eq}[1]{Eq.~(\ref{#1})}
\newcommand{\Eq}[1]{Equation~(\ref{#1})}
\newcommand{\tab}[1]{Table~\ref{#1}}

\renewcommand{\vec}[1]{\ensuremath\boldsymbol{#1}}
\renewcommand{\epsilon}[0]{\varepsilon}

\newcommand{\myscale}{0.62}

\begin{document}


\title{
  Efficacy of the DFT+$U$ formalism for modeling hole polarons in perovskite oxides
}

\author{Paul Erhart}
\email{erhart@chalmers.se}
\affiliation{
  Chalmers University of Technology,
  Department of Applied Physics,
  S-412 96 Gothenburg, Sweden
}
\author{Andreas Klein}
\affiliation{
  Technische Universit\"at Darmstadt,
  Institut f\"ur Materialwissenschaft,
  64287 Darmstadt, Germany
}
\author{Daniel {\AA}berg}
\author{Babak Sadigh}
\affiliation{
  Lawrence Livermore National Laboratory,
  Chemistry, Materials and Life Sciences Directorate,
  Livermore, 94550, California, USA
}

\begin{abstract}
We investigate the formation of self-trapped holes (STH) in three prototypical perovskites (\ST, \BT, \PT) using a combination of density functional theory (DFT) calculations with local potentials and hybrid functionals. First we construct a local correction potential for polaronic configurations in \ST\ that is applied via the DFT+$U$ method and matches the forces from hybrid calculations. We then use the DFT+$U$ potential to search the configuration space and locate the lowest energy STH configuration. It is demonstrated that both the DFT+$U$ potential and the hybrid functional yield a piece-wise linear dependence of the total energy on the occupation of the STH level suggesting that self-interaction effects have been properly removed. The DFT+$U$ model is found to be transferable to \BT\ and \PT, and STH formation energies from DFT+$U$ and hybrid calculations are in close agreement for all three materials. STH formation is found to be energetically favorable in \ST\ and \BT\ but not in \PT, which can be rationalized by considering the alignment of the valence band edges on an absolute energy scale. In the case of \PT\ the strong coupling between Pb $6s$ and O $2p$ states lifts the valence band minimum (VBM) compared to \ST\ and \BT. This reduces the separation between VBM and STH level and renders the STH configuration metastable with respect to delocalization (band hole state). We expect that the present approach can be adapted to study STH formation also oxides with different crystal structures and chemical composition.
\end{abstract}

\maketitle

\section{Introduction}

In materials with at least partially ionic bonding character, most notably halides and oxides, charge excitations can couple to lattice modes leading to the formation of polarons \cite{Sch06, StoGavShl07, ShlMcKSus09}. In the limit where the coupling between charge excitations and phonons is strong one obtains so-called small or Holstein polarons \cite{Hol59a}, which are characterized by very large but localized lattice distortions. Their motion can be considered classical \cite{ShlSto93, StoGavShl07} and typically exhibits an exponential temperature dependence. As a result the presence of polarons usually implies low mobilities, which are detrimental for many applications.

Perovskite oxides exhibit a broad variety of interesting phenomena including but not limited to ferroelectricity, multiferroicity, strong correlation, and low-dimensional electron gases. Applications are abundant as well as they are being used for example in electronics, catalysis, and thermoelectrics. Polaronic effects in these materials have been discussed for some time \cite{Ihr76, Sch06} and have also been investigated theoretically \cite{StaPin00, PinSta02, QiuWuNas05, QiuJiaTon08}. While the existence of ``bound'' small hole polarons \footnote{In particular in the titanates one can also observe electron polarons, which are associated with the reduction of Ti$^{4+}$ to Ti$^{3+}$.}, which are associated for example with acceptor defects in oxides, is well established the presence of ``free'' (or self-trapped) polarons in oxides, which are naturally also much harder to observe, has been doubted \cite{Sch06}. By contrast earlier computational studies did identify polarons in \BT\ based on embedded cluster Hartree-Fock (HF) calculations \cite{StaPin00, PinSta02} and \ST\ based on a suitably parametrized many-body model Hamiltonian \cite{QiuWuNas05, QiuJiaTon08}. This provides the motivation for the present work, in which we explore self-trapped hole (STH) polarons in three prototypical oxidic perovskites, \ST, \BT, and \PT. As will be discussed in detail below the description of polaronic effects is sensitive to the level of theory that is being employed. We therefore compare different techniques and carefully assess their respective predictiveness. We find that self trapped polarons are energetically favorable in both \ST\ and \BT\ but not in \PT. The formation energies for small polarons in the first two materials are $-0.1\,\eV$ and $-0.2\,\eV$, respectively, while the associated lattice distortions are less than 0.12\,\AA\ for individual atoms. The lack of spontaneous STH formation in \PT\ can be traced to the strong coupling between Pb $6s$ and O $2p$ states, which raises the valence band maximum. The observed correlation between STH formation energies and VBM position suggests a simple approximate predictor for STH formation in this class of materials.

In the following section we discuss shortcomings of common electronic structure methods such as density functional theory (DFT) and HF as well as related approaches with respect to the description of polarons. In particular we argue for the suitability of the DFT+$U$ method for describing the systems of interest in this work. In \sect{sect:results_forces} it is shown that DFT+$U$ functionals can be parametrized to reproduce the forces obtained from higher level (and computationally much more expensive) calculations based on hybrid exchange-correlation (XC) functionals. Using a suitable parametrization we then explore in \sect{sect:results_sth_structure} the STH configuration space in the case of \ST, identify as well as characterize the groundstate configuration, and also verify the results by comparison with hybrid XC functional calculations. In \sect{sect:results_linearity} we then demonstrate that both DFT+$U$ and hybrid XC calculations yield piece-wise linear behavior for the total energy as a function of fractional charge, which demonstrates their suitability for the present purpose and confounds their predictiveness. Finally, \sect{sect:results_BT_PT} presents results for \BT\ and \PT, which are rationalized in terms of the valence band alignment between the different materials.

\section{Methodology}
\label{sect:methodology}

Density functional theory (DFT) based on semilocal XC functionals is a powerful tool for electronic structure calculations. It fails, however, to reproduce polaron formation in many condensed systems not only quantitatively but qualitatively \cite{GavSusShl03, SadErhAbe14}. This shortcoming can be traced to the self-interaction (SI) intrinsic to common semilocal approximations to the XC functional including the local density approximation (LDA) as well as the generalized gradient approximation (GGA). The importance of SI corrections in calculations based on semilocal XC functionals was first discussed by Perdew and Zunger \cite{PerZun81}, who also proposed a correction scheme that works well for atoms but falls short when it comes to molecules or solids. The issue has been addressed more recently in terms of variation of the total energy with respect to fractional changes in the electronic occupations \cite{CocGir05, DabFerPoi10, LanZun09, ZawRosWed11}. An exact functional should lead to piece-wise linear behavior with discontinuous derivatives at integer occupations \cite{DabFerPoi10}. DFT calculations based on semilocal XC functionals deviate from this requirement and overbind (concave dependence), whereas HF calculations tend to underbind (convex dependence) \cite{PerRuzCso07, MorCohYan09}. These opposite behaviors can be exploited in parametrizations of hybrid functionals that minimize the deviation from piece-wise linearity \cite{LanZun09}. 

The overbinding of LDA/GGA and associated lack of localization is also at the heart of another DFT failure that is associated with partially occupied $d$-states in transition metal oxides, which are erroneously predicted to be metallic. The DFT+$U$ (originally LDA+$U$) method \cite{AniZaaAnd91}, which was proposed to overcome this error, adds a Hubbard-like term to the total energy functional and solves it using the self-consistent field approach in the independent particle approximation. Hence, the standard DFT$+U$ approach treats the Hubbard term within the mean-field approximation. Consequently, this method effectively adds a Fock-exchange term inside each atomic sphere to the standard LDA/GGA functionals.
\begin{align}
  E_\text{DFT+$U$} &= E_\text{DFT} + \frac{1}{2} \sum_{I\alpha} U_\alpha n_{I,\alpha} \left( 1 - n_{I,\alpha} \right).
\end{align}
Above the summation runs over sites $I$ and projection channels $\alpha$, and $n_{I\alpha}$ is the occupancy of chancel $\alpha$ at site $I$. The resulting correction is quadratic in the parameter $U$ and it was argued that self-consistent $U$ parameters should correspond to piece-wise linear behavior \cite{CocGir05}. The DFT+$U$ approach was originally derived to account for the extra on-site Coulomb repulsion that occurs in strongly correlated systems due to the coexistence of atomic-like $d$ or $f$-electron states with delocalized band states. However, considering the mathematical similarity of the DFT+$U$ method to hybrid DFT techniques, it should in principle also be appropriate for treating atomic-like polaronic states in normal insulators, which often possess $p$-character, no matter whether they happen to be bound to point defects such as vacancies or have been self-trapped by lattice distortions. The hole states derived from O $2p$ orbitals in wide band gap oxides constitute a large class of problems that fall into this category. 

In order to establish a more detailed understanding of the applicability of the DFT+$U$ method to the problem of polarons in insulators, we discuss here polarons in two classes of wide band gap insulators: halides and functional oxides. In halides the formation of an STH \nocite{Kan55} \footnote{In halides STH configurations are for historical reasons commonly referred to as $V_K$-centers, Ref.~\onlinecite{Kan55}.} occurs by dimerization of two halogen ions, which can be formally described as the reaction $X^- + X^- \rightarrow X_2^- + e'$, where $X$ represents the halogen and $e'$ denotes the band electron that is created in the process. The resulting hole is localized {\em between} the two halogens. By contrast in oxides hole self-trapping is a single-center process \cite{Sch06} corresponding to the reduction of an oxygen ion $\text{O}^{2-} \rightarrow \text{O}^- + e'$. Let us now consider a charge balanced oxide crystal comprising $N$ oxygen atoms, from which one electron is removed. STH formation implies $N\text{O}^{2-} \rightarrow (N-1)\text{O}^{2-} + \text{O}^{-} + e'$ whereas LDA/GGA yields $N\text{O}^{2-} \rightarrow N\text{O}^{2- + 1/N}  + e'$. This is related to the lack of charge disproportionation in partially filled $d$-states for e.g., for ions in solution \cite{SitCocMar06} and similarly originates from the overbinding/lack of localization in LDA/GGA alluded to above. This motivates the addition of a DFT+$U$ term to the total energy functional where the penalty term is applied to the O $2p$ states. This approach is similar to the potential operator defined in Ref.~\onlinecite{LanZun09}.

In the present work we employ hybrid functionals as well as the DFT+$U$ method, and check their predictions with regard to piece-wise linearity. It should be pointed out that hybrid functionals are more generally applicable but at the same time computationally much more expensive than DFT+$U$ calculations. All calculations were carried out using the project augmented wave method \cite{Blo94, *KreJou99} as implemented in the Vienna ab-initio simulation package \cite{KreHaf93, *KreHaf94, *KreFur96a, *KreFur96b} using a plane wave energy cutoff of 500\,eV for volume relaxations and 400\,eV otherwise. We used the LDA functional in combination with the DFT+$U$ method. Hybrid DFT calculations were performed using the range-separated HSE06 functional \cite{HeyScuErn03}.

\section{Results}

\begin{figure*}
  \centering
\includegraphics[scale=\myscale]{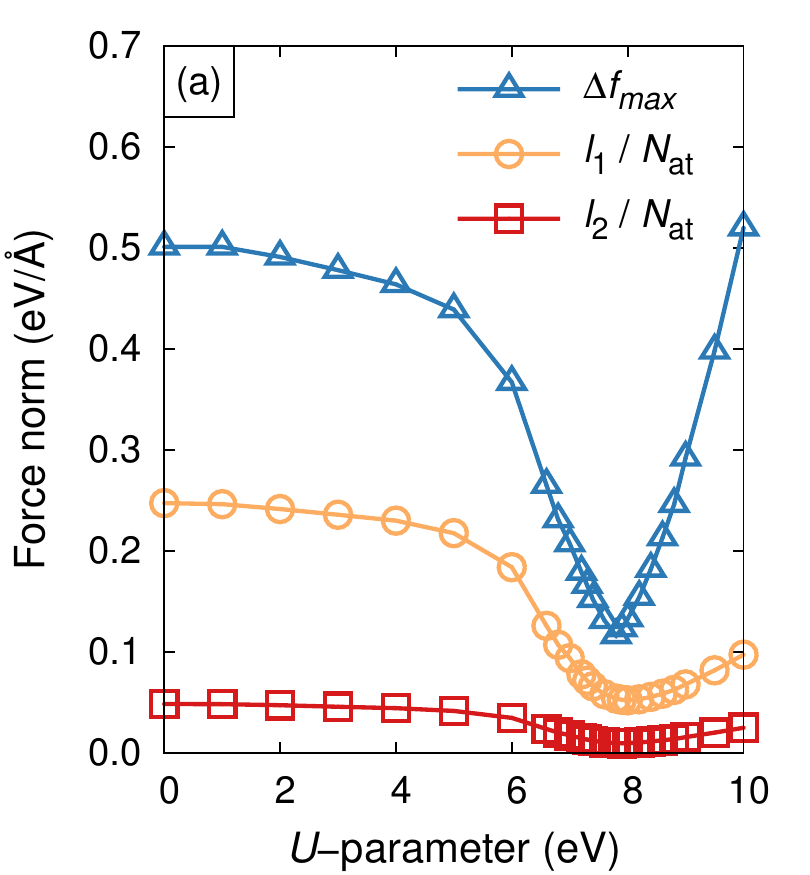}
\includegraphics[scale=\myscale]{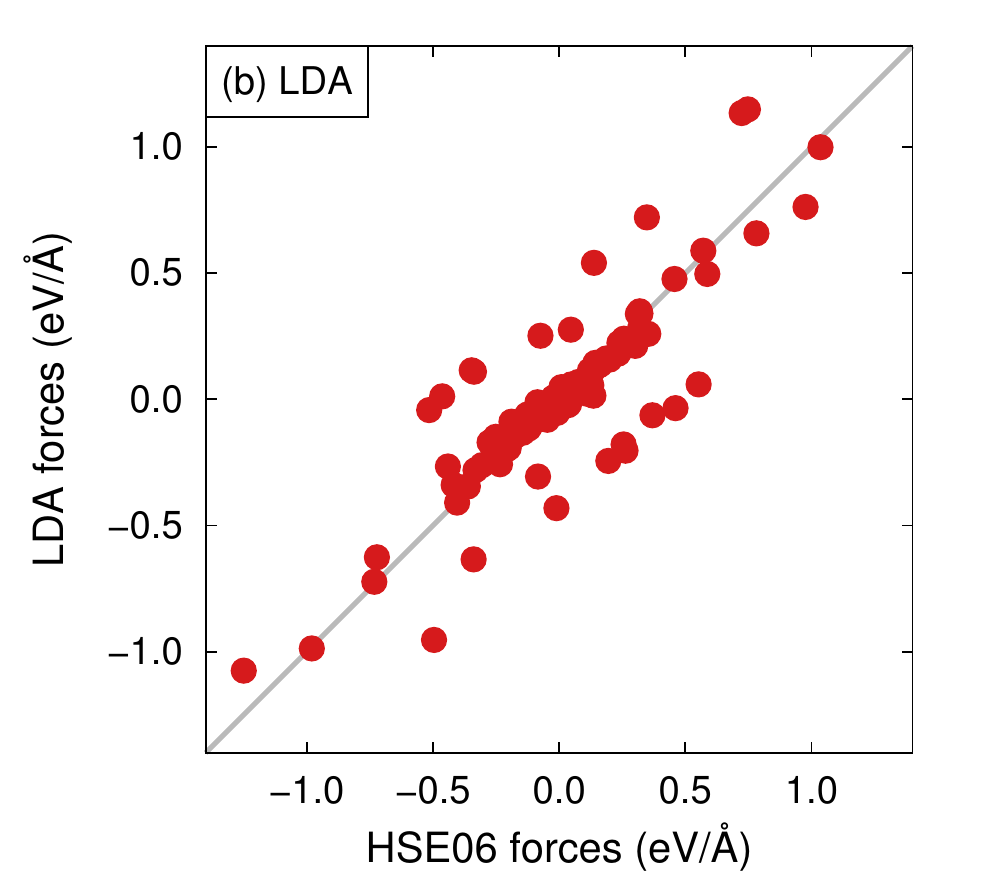}
\includegraphics[scale=\myscale]{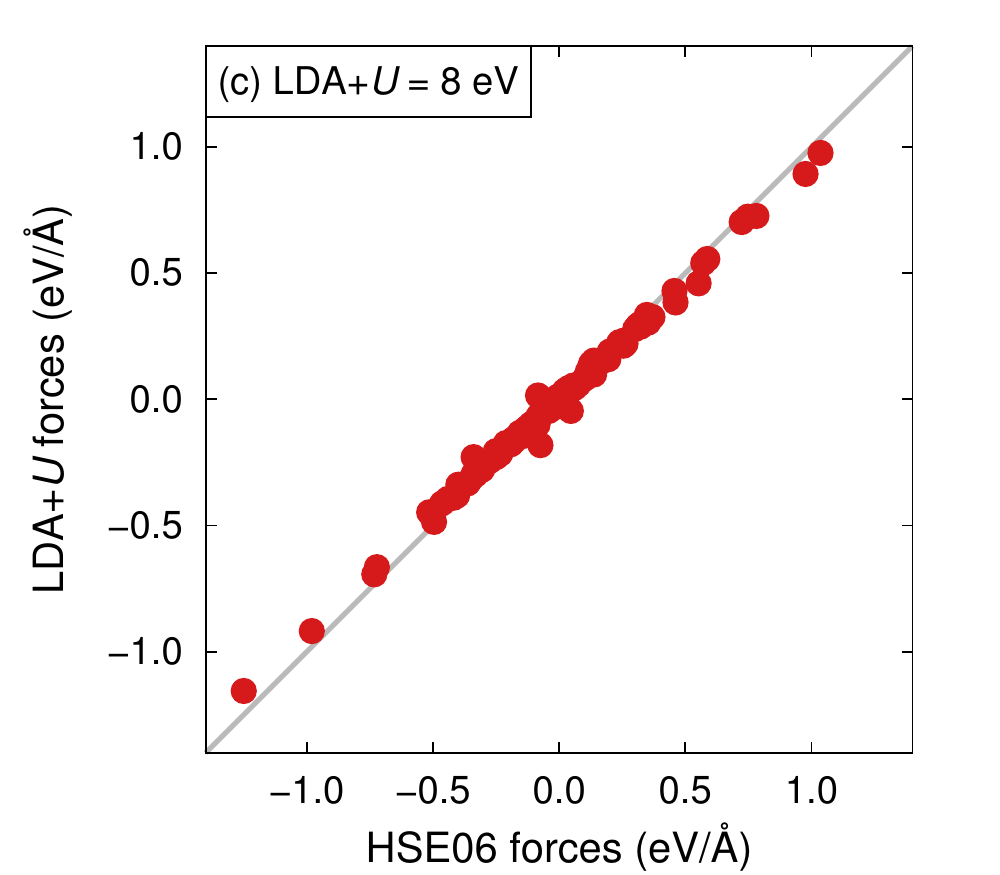}
  \caption{
    (a) Force matching between LDA+$U$ models with $U$ applied to O $2p$ states and the HSE06 hybrid functional for cubic \ST\ based on $l_1$ and $l_2$-norm as well as maximum deviation of forces $\Delta f_{max}$ (40-atom cell, one hole).
    Comparison of force components from (b) LDA ($U=0\,\eV$) and (c) LDA+$U$ with $U=8\,\eV$ with HSE06 calculations.
  }
  \label{fig:forces}
\end{figure*}

\subsection{LDA+$U$ model for sampling the energy landscape}
\label{sect:results_forces}

Let us first consider STH formation in the cubic phase of \ST, and determine the associated lattice distortion.
While an {\it a-priori} assumption regarding the geometric structure or symmetry of the ionic displacements belong to the STH should be avoided an exhaustive search over configurations using a hybrid functional is computationally exceedingly prohibitive. LDA/GGA calculations on the other hand fail to produce STH formation all together, which is not surprising based on previous experience with STH formation in halides \cite{GavSusShl03}. In fact the forces computed within the LDA differ substantially from those obtained from HSE06 calculations as evident from the left most data points in \fig{fig:forces}(a) as well as \fig{fig:forces}(b). These forces were computed for a $2\times2\times2$ \ST\ supercell with one electron removed from the system and random displacements drawn from a Gaussian distribution with a standard deviation of 0.02\,\AA\ using a $2\times2\times2$ Monkhorst-Pack mesh for sampling the Brillouin zone.

Based on the arguments provided in the previous section there should, however, exist a DFT+$U$ parametrization that yields an energy landscape in closer agreement with the HSE06 data. This is indeed the case as illustrated in \fig{fig:forces}(a), which shows the maximum difference between equivalent atomic forces from HSE06 and LDA+$U$ calculations, $\Delta f_{max} = \max_i\left\{\left|\vec{f}_i^{\text{LDA}+U}-\vec{f}_i^{\text{HSE06}}\right|\right\}$ along with the $l_1$ and $l_2$-norms of the force difference vector. Here the LDA+$U$ potential is applied to the O $2p$ state. Very close agreement is obtained for $U\approx8\,\eV$ as further illustrated by direct comparison of all force components in \fig{fig:forces}(c). We also considered random displacements with larger amplitudes corresponding to standard deviations of 0.05\,\AA\ and 0.10\,\AA, which resulted in maximum forces in excess of 3 and 11\,eV/\AA, respectively, and observed similarly good agreement.

\subsection{Geometric and electronic structure of STH in \ST}
\label{sect:results_sth_structure}

\begin{figure}
\includegraphics[width=0.41\linewidth]{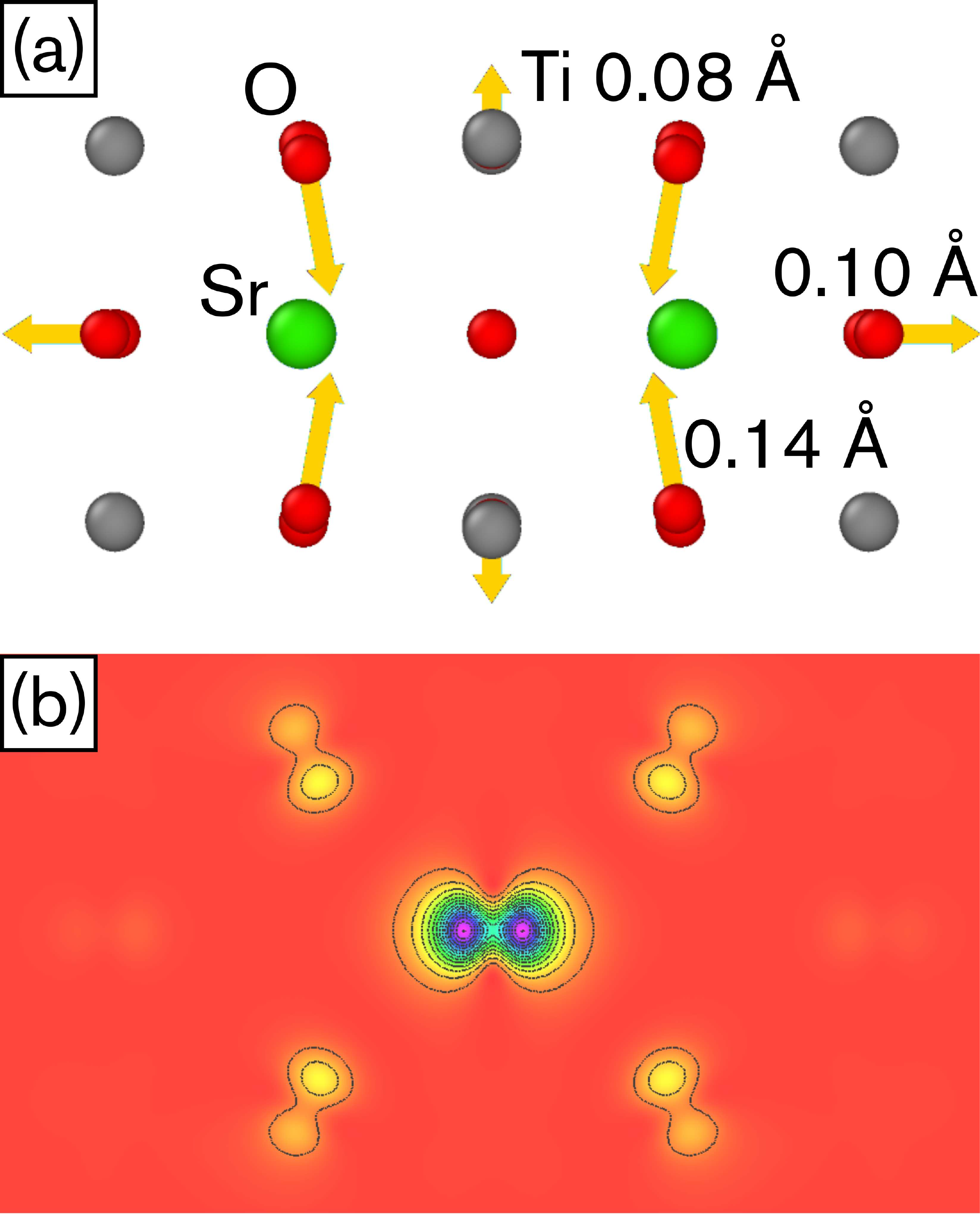}
\includegraphics[scale=\myscale]{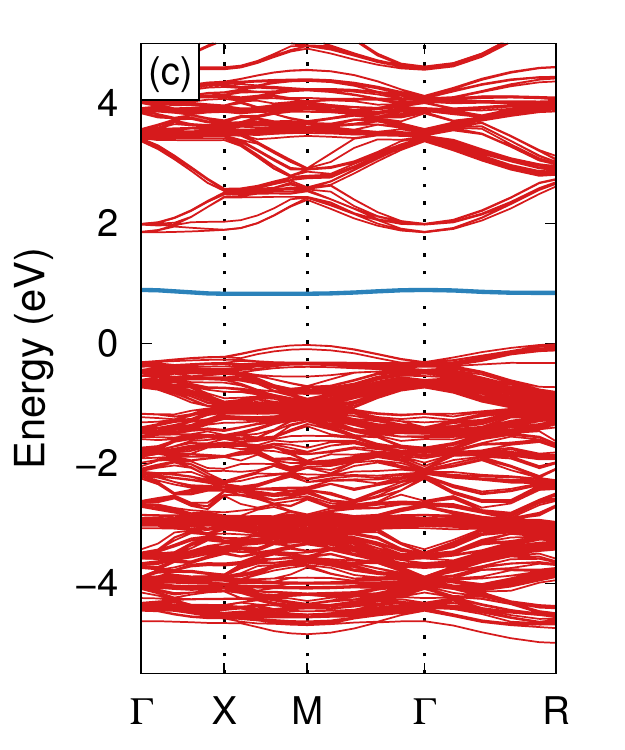}
  \caption{
    (a) Geometric structure of STH in cubic \ST\ and (b) hole density corresponding to STH level projected onto (100) plane.
    (c) Band structure for $3\times3\times3$ supercell showing the STH level in the band gap (LDA+$U$).
    $\vec{k}$-point labels are equivalent to the primitive cell.
  }
  \label{fig:sth}
\end{figure}

This type of LDA+$U$ calculations for the systems of interest in this work are more than two orders of magnitude faster than equivalent hybrid calculations, and therefore enabled us to carry out a systematic search for minima in the potential energy landscape, which yielded one symmetrically distinct configuration. To this end we used $3\times3\times3$ supercells and a $2\times2\times2$ Monkhorst-Pack mesh. For the final configuration calculations were also carried out using the HSE06 functional. Already for the LDA+$U$ configuration the maximum force was less than 0.2\,eV/\AA. The structure was then relaxed until the maximum force fell below 20\,meV/\AA, which changed the energy by less than 0.01\,eV and positions by less than 0.03\,\AA. This demonstrates that in terms of geometries our LDA+$U$ parametrization works remarkably well.

The STH configuration obtained in this way is shown in \fig{fig:sth}(a). STH formation leads to the emergence of a localized level that as shown in \fig{fig:sth}(b,c) is located at a single oxygen site and exhibits O $2p$ character. While the oxygen atom at the center of the STH configuration remains at its ideal lattice site the two nearest Ti neighbors are displaced outward by 0.09\,\AA\ along $\left<100\right>$ and the four next-nearest O neighbors move inward by 0.12\,\AA\ along $\left<110\right>$. Note that overall the ionic relaxations are small, while the hole has been completely localized. This is rather different from most $V_K$-centers in halides, which typically exhibit lattice distortions on the order of 0.5\,\AA\ and above \cite{SadErhAbe14}. As a result the migration of the STH involves small changes in the ion positions when the localized charge is transferred between nearest-neighbor oxygen sites. This in turn implies that the non-adiabatic, in particular diabatic, \cite{Mar56} potential-energy landscapes must be calculated in order to obtain correct rates of diffusion of these species.

\subsection{Piece-wise linearity}
\label{sect:results_linearity}

\begin{figure}
\includegraphics[scale=\myscale]{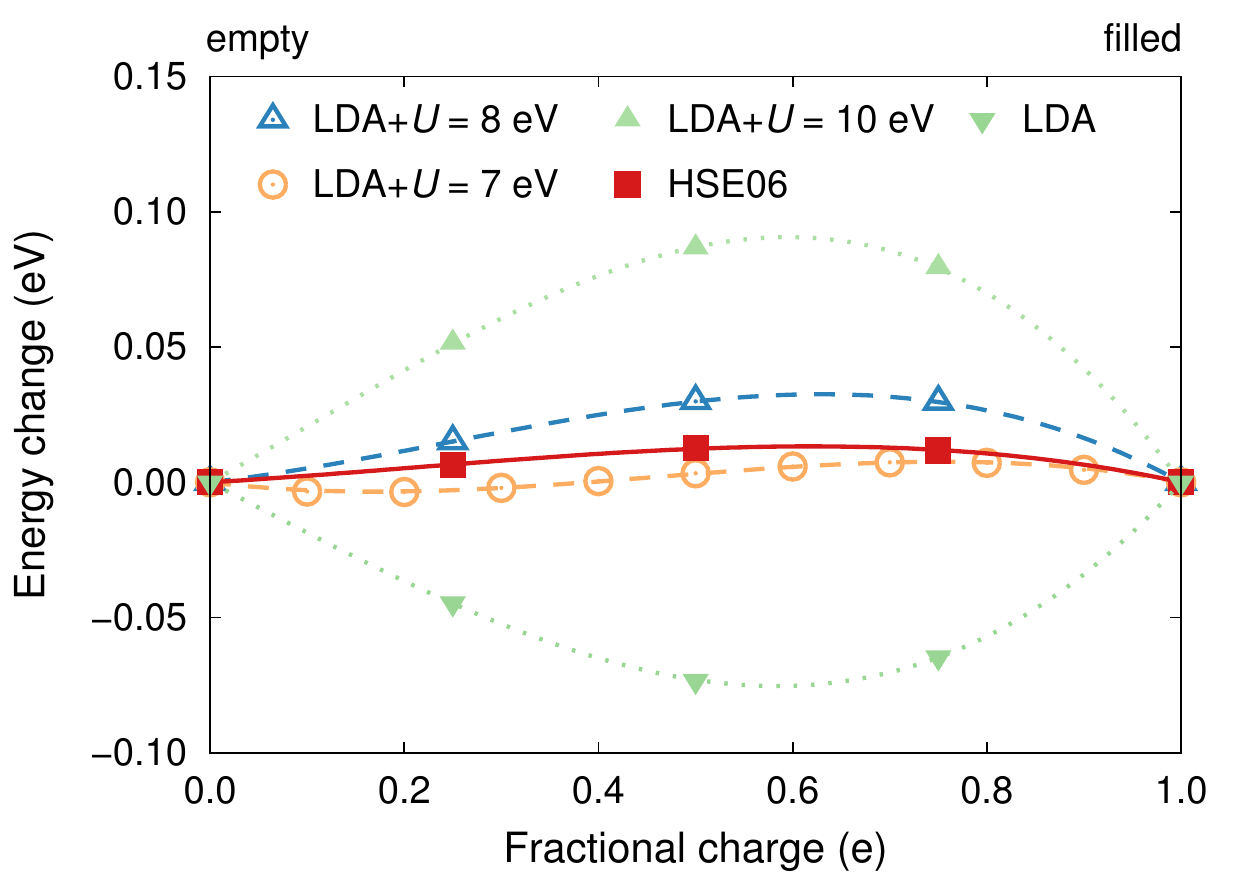}
  \caption{
    Variation in total energy with excess charge calculated using the HSE06 hybrid functional as well as several LDA+$U$ models. Both HSE06 and $U=7\,\eV$ yield close to piece-wise linear behavior.
  }
  \label{fig:fracocc}
\end{figure}

Thanks to the emergence of a localized level we now have a system, for which we can explicitly study the variation of the total energy with fractional occupation as the localized level is gradually being filled. \Fig{fig:fracocc} shows that the HSE06 functional actually performs very well in producing a piece-wise linear variation of the energy between integer occupations. At the same time HSE06 predicts a band gap of 3.1\,eV in good agreement with the experimental value of 3.25\,eV \cite{BenElsFre01} (compare \tab{tab:results}). This demonstrates that in the present case HSE06 {\em simultaneously} provides a good description of the ideal and the self-trapped structure, confounding the suitability of this hybrid functional for the present purpose. Next we consider the variation of the total energy with fractional occupation for a series of LDA+$U$ functionals with varying $U$. With increasing $U$ the behavior changes from concave to convex as anticipated in \sect{sect:methodology}.
\footnote{
  One should note that image charge interactions also lead to quadratic term in fractional occupation. The very large dielectric constants of the oxides considered in this study, however, imply that in the present cases this effect is negligibly small. This is also confirmed by test calculations using larger supercells. Note that a self-consistent determination of the $U$ parameter following for example the procedure outlined in Refs.~\onlinecite{AniGun91,CocGir05} avoids this ambiguity and is therefore suitable for arbitrary oxides.
}
The best agreement is obtained for a $U$-parameter of 7\,eV, which is close to the value of 8\,eV obtained by matching the forces and demonstrates the consistency of our approach. For simplicity in the following, however, we continue using a value of 8\,eV. The effect of this choice on the results is small.

%%% TEXEXPAND: INCLUDED FILE MARKER ./tab_results.tex
\begin{table}
\centering
\caption{
  Compilation of properties computed for \ST, \PT, and \BT\ using different approximations to the XC functional.
  In the case of the LDA+$U$ functional a $U$ parameter of 8\,eV was applied exclusively to the O $2p$ states.
  All STH calculations were carried out at the respective equilibrium lattice parameters using $3\times3\times3$ supercells (135 atoms) and $2\times2\times2$ Monkhorst-Pack $k$-point grids with the exception of the calculation of the position of the STH level, which was carried out using $5\times5\times5$ $\Gamma$-centered grids.
}
\label{tab:results}
\newcommand{\spr}[1]{\multicolumn{1}{c}{#1}}
\newcommand{\sprfull}[1]{\multicolumn{5}{l}{#1}}
\begin{tabularx}{0.95\columnwidth}{p{1mm}X*{3}d}
\hline\hline
& Method & \spr{\ST} & \spr{\PT} & \spr{\BT} \\
\hline
\sprfull{Lattice constant (\AA)} \\
%& PBE        & 3.949 & 3.965 & 4.036 \\
& LDA        & 3.871 & 3.888 & 3.957 \\
& LDA+$U$    & 3.850 & 3.868 & 3.938 \\
& HSE06      & 3.908 & 3.923 & 3.993 \\[6pt]

\sprfull{Band gap (eV)} \\
%& PBE        &  1.66  &  1.55  &  1.56  \\
& LDA        &  1.70  &  1.41  &  1.61  \\
& LDA+$U$    &  1.85  &  1.37  &  1.78  \\
& HSE06      &  3.09  &  2.54  &  2.94  \\
& Experiment (Refs.~\onlinecite{BenElsFre01, SchLiChe11, *LiMorKle13}) $^*$ &   3.25    &  (3.2)   &  (3.2)   \\[6pt]

%%\sprfull{Dielectric constant, electronic contribution} \\
%%& LDA        &        &   7.18 &   8.98 \\
%%& LDA+$U$    &   6.33 &   7.05 &   8.45 \\
%%& HSE06      &        &        &        \\
%%& Experiment &        &        &        \\[6pt]
%%
%%\sprfull{Dielectric constant, ionic contribution} \\
%%& LDA        &        &        &  14.54 \\
%%& LDA+$U$    & 230.39 & 763.31 &  14.29 \\
%%& HSE06      &        &        &        \\
%%& Experiment &        &        &        \\[6pt]

\sprfull{Formation energy $\Delta E_f$ according to \eq{eq:eform} (eV)} \\
& LDA+$U$ &  -0.09  &  +0.24  &  -0.20  \\
& HSE06   &  -0.09  &  +0.32  &  -0.25  \\[6pt]

\sprfull{STH level relative to VBM [compare \fig{fig:sth}(c)] (eV)} \\
& LDA+$U$ &  0.88  &  0.56  &  0.87  \\[6pt]

\sprfull{Valence band offset relative to \BT\ [compare \fig{fig:banddos}(a)] (eV)} \\
& LDA     & 0.2 & 0.8 \\
& LDA+$U$ & 0.1 & 0.9 \\
& HSE06   & 0.1 & 0.9 \\
& Experiment (Ref.~\onlinecite{SchLiChe11, *LiMorKle13}) & 0.0 & 1.2 \\[3pt]
\hline\hline
\end{tabularx}
\\
\noindent
\begin{flushleft}
$^*$ The experimental data are room temperature values corresponding in the case of \BT\ and \PT\ to the tetragonal phase.
\end{flushleft}
\end{table}
%%% TEXEXPAND: END FILE ./tab_results.tex

\subsection{STH formation energies}

\begin{figure}
\includegraphics[scale=\myscale]{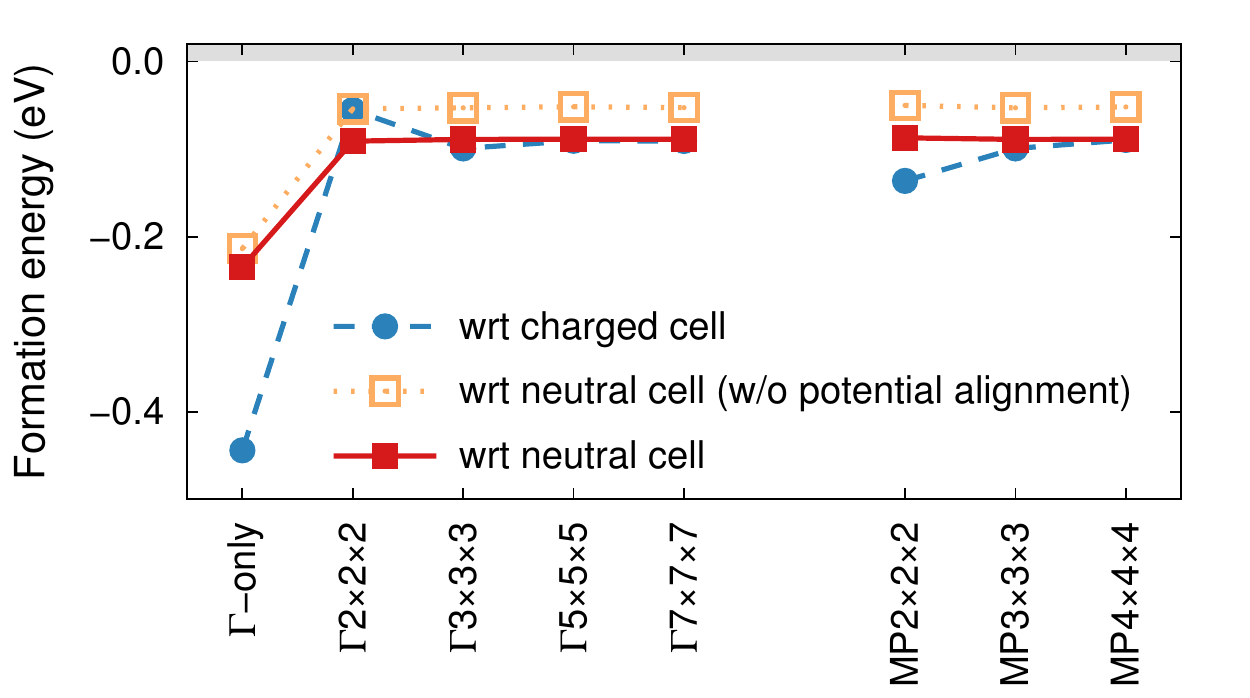}
  \caption{
    Convergence of STH formation energy computed via Eqs.~(\ref{eq:eform}) and (\ref{eq:eform_q1}) with $\vec{k}$-point grid.
    The difference between the open and closed squares illustrates the effect of the potential alignment correction term $\Delta v_\text{PA}$.
  }
  \label{fig:eform_kpts}
\end{figure}

The STH formation energy is given by
\begin{align}
  \Delta E_f &= E_\text{STH} - E_{id}^0 + q \left( \epsilon_\text{VBM} + \Delta v_\text{PA}(q) \right),
  \label{eq:eform}
\end{align}
where $E_\text{STH}$ and $E_{id}^0$ are the total energy of the STH configuration and the corresponding charge neutral ideal configuration, $q$ denotes the excess charge density from the STHs in each supercell, $\epsilon_\text{VBM}$ is the valence band maximum (VBM) in the perfect crystal, and finally $\Delta v_\text{PA}(q)$ denotes the potential alignment (PA) correction, which originates from the fact that the total energy per unit cell of an infinite system subject to periodic boundary conditions is only known to within a constant due to the exact cancellation of three infinities: the electron-electron, the ion-ion and the electron-ion electrostatic energies. The value of this constant can shift when the overall electronic charge state of each supercell in a periodic system is changed, while the total energy per unit cell is kept finite by adding a compensating homogeneous background charge \cite{LanZun08, KomRanPas12}. In the present work $\Delta v_\text{PA}$ has been obtained as the difference in the electrostatic potential at the ionic cores between the charged and the neutral ideal cells, $\Delta v_\text{PA} = v_{id}(q) - v_{id}(0)$. Note that since the contribution to the total energy $q\times\Delta v_\text{PA}(q)$ is caused by purely electrostatic interactions due to the delocalized electron/hole carrier density $q$ added to each supercell and the compensating homogeneous background charge, to lowest order it has to scale quadratically with the excess charge density $q$, or in other words the density of STH. Hence to lowest order 
\begin{align}
  \Delta v_\text{PA}(q) = q \Delta v'_\text{PA}  + \mathcal{O}(q^2).
\end{align}

\Eq{eq:eform} is equivalent to the formation energy expression obtained in the case of point defects \cite{ZhaNor91, AbeErhWil08}, where the latter contains a term that accounts for changes in the number of atoms, which is of no concern here. 

\begin{figure*}
\includegraphics[scale=\myscale]{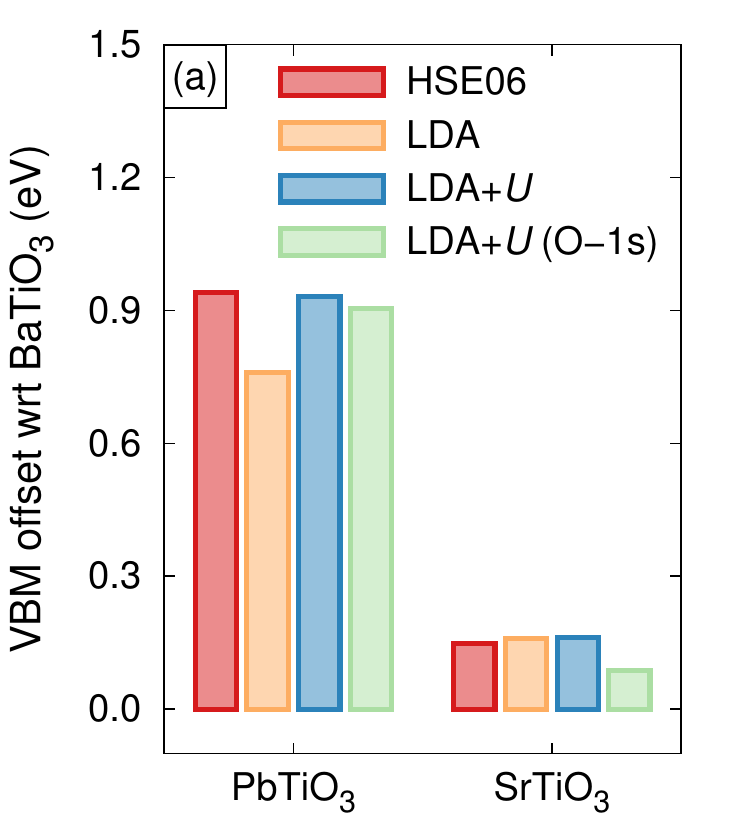}
\includegraphics[scale=\myscale]{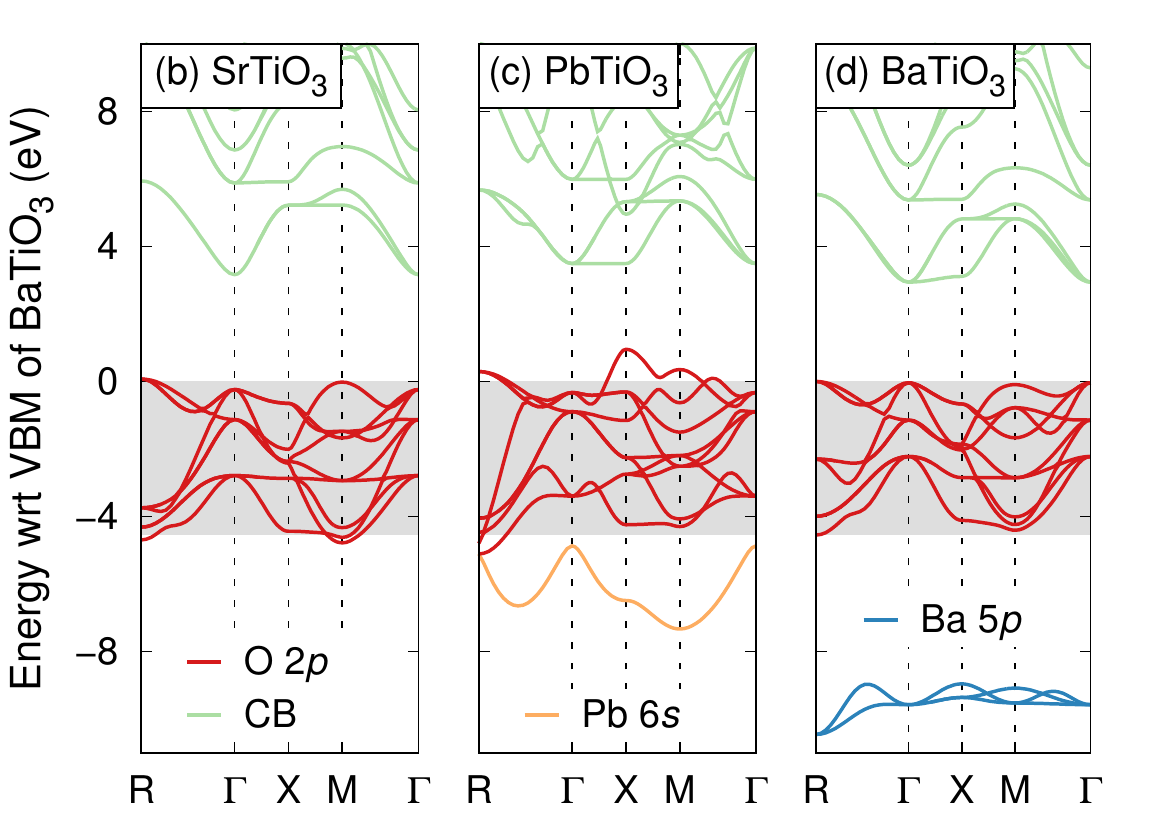}
\includegraphics[scale=\myscale]{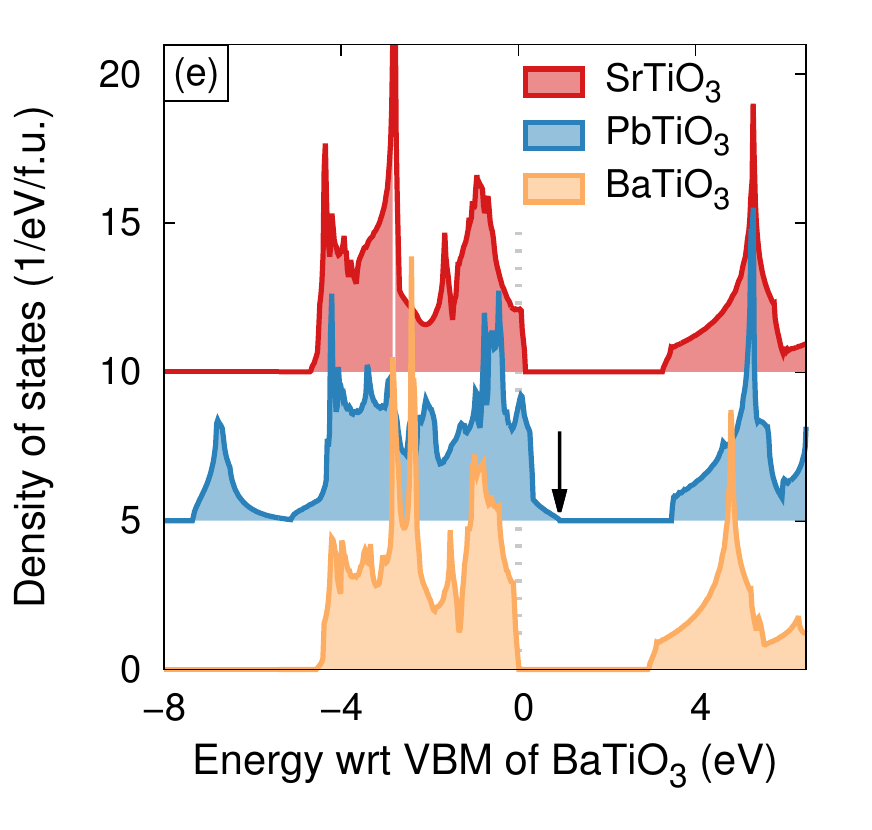}
  \caption{
    (a) Band line up between \BT, \PT, and \ST\ according to different XC functionals.
    (b-d) Band structures and (e) densities of states for \ST, \PT, and \BT\ according to LDA+$U$ calculations with scissor corrections. The significantly higher VBM in \PT\ originates from the coupling between O $2p$ and Pb $6s$ states. The latter appear just below the O $2p$ derived topmost valence band and are visible both in (c) and (e) between approximately $-4$ and $-8\,\eV$.
    The bands in (b-d) have been colored according to the dominant character of the entire band manifold to indicate the assignments qualitatively rather than quantitatively.
    The arrow in (e) indicates the position of the VBM in \PT.
  }
  \label{fig:banddos}
\end{figure*}

One should observe that the valence band maximum $\epsilon_\text{VBM}$, which here is taken as the highest occupied Kohn-Sham eigenvalue, is also related to the total energy of the ideal cell according to
\begin{align}
  \epsilon_\text{VBM}
  &= \lim_{\delta q\rightarrow 0} \left[ E_{id}^{\delta q} - E_{id}^0 \right] / \delta q - \Delta v'_\text{PA} ,
\end{align}
where $\delta q$ is a small hole charge and $E_{id}^{\delta q}$ refers to the total energy of an ideal cell with an excess hole carrier in the infinite dilution limit. This can be compared to the following equivalent expression for the STH formation energy
\begin{align}
  \Delta \widetilde{E_f} &= E_\text{STH} - E_{id}^{+1}.
  \label{eq:eform_q1}
\end{align}

When computing formation energies of charged defects one usually also must account for image charge interactions corresponding to the spurious binding between an array of localized charges and a homogeneous background \cite{MakPay95, LanZun08, KomRanPas12}. In the present case strong dielectric screening renders this interaction, however, exceedingly small. A very conservative estimate based on the monopole-monopole interaction, which provides an upper bound for this contribution, yields a value of 7\,meV, which due to its smallness has not been considered further.

We have performed a $\vec{k}$-point convergence study of \eq{eq:eform} and \eq{eq:eform_q1} as shown in \fig{fig:eform_kpts}.\footnote{The total energies of the STH and the neutral ideal cell converge quickly with respect to the $\vec{k}$-point mesh whereas the charged ideal cell requires more $\vec{k}$-points for convergence due to partially occupied states near the VBM.} The figure also illustrates the importance of PA corrections and the fact that for sufficiently dense sampling of the Brillouin zone both \eq{eq:eform} and \eq{eq:eform_q1} converge to the same value.

The formation energies computed using LDA+$U$ with $U(\text{O}~2p)=8\,\eV$ for a $2\times2\times2$ Monkhorst-Pack mesh are $\Delta E_f=-0.09\,\eV$ and $\Delta\widetilde{E_f}=-0.14\,\eV$, in good agreement with the values of $\Delta E_f=-0.09\,\eV$ and $\Delta\widetilde{E_f}=-0.16\,\eV$ obtained using the HSE06 hybrid functional, where the latter calculations required more than two orders of magnitude more computer time.
We can compare these values with results from a model Hamiltonian approach \cite{QiuWuNas05, QiuJiaTon08}, which was parametrized using primarily experimental data. The authors obtained an energy gain of $-0.2\,\eV$ upon STH formation, which is in reasonable agreement with the present data.

\subsection{Extension to \BT\ and \PT}
\label{sect:results_BT_PT}

In addition to \ST\ we have considered two other cubic perovskitic titanates, namely \BT\  and \PT. For \BT\ our calculations predict STH formation with a formation energy of $\Delta E_f=-0.20\,\eV$ at the LDA+$U$ level (with $U=8\,\eV$ applied to O $2p$ states as in the case of \ST\ discussed above) and $-0.25\,\eV$ at the HSE06 level (compare \tab{tab:results}). In the case of \PT\ we obtain formation energies of $\Delta E_f=+0.24\,\eV$ and $+0.32\,\eV$ from LDA+$U$ and HSE06 calculations, respectively. Note that the good agreement between LDA+$U$ and HSE06 was obtained by simply using the same $U$ parameter as in the case of \ST, which suggests that the parameter is reasonably transferable at least within this group of rather similar oxides.

It is instructive to compare our results with previous studies. Embedded cluster calculations based on the HF approximation yielded ``hole trapping energies'', which should be comparable to the formation energies obtained in the present study, of $-1.49\,\eV$ for cubic \BT\ (Ref.~\onlinecite{StaPin00}) and $-0.87\,\eV$ for tetragonal \BT\ (Ref.~\onlinecite{PinSta02}). Given the tendency of the HF method to overestimate localization as discussed in \sect{sect:methodology} it is not surprising that these values are much more negative than the ones obtained in the present work. They are also associated with band gaps of 5.4 and 6.1\,eV for cubic and tetragonal \BT, respectively, that are considerably larger than the experimental values (see \tab{tab:results}). In agreement with the present study the relaxation pattern obtained in these studies includes an outward relaxation of the two nearest Ti neighbors and an inward relaxation of the next nearest O neighbors by about 0.1\,\AA.

The positive STH formation energy in the case of \PT\ implies that localized holes are metastable and energetically less favorable than their delocalized (band) counterparts and thus should not occur under normal conditions. This begs the question what distinguishes \PT\ from \BT\ and \ST\ when it comes to STH formation. The situation can be rationalized by considering the position of the VBM in the different materials. \footnote{Note that a similar form of alignment was observed in the case of rutile and anatase TiO$_2$, see Ref.~\onlinecite{DeaAraFra12}.} \Fig{fig:banddos}(a) shows the shift in the VBM for \PT\ and \ST\ with respect to \BT, which reveals that with respect to an absolute energy scale the VBM in \PT\ is markedly higher than in either \BT\ or \ST. The energy scales were aligned using the core potential at the oxygen sites. Alternatively one can also employ the O $1s$ levels for alignment, which as demonstrated by the last data set in \fig{fig:banddos}(a), leads to very similar results. This approach was employed previously to determine the VBM offset between the rutile and anatase phases of titania \cite{PfeErhLi13}. Also the offsets thus obtained are in good agreement with recent experimental data \cite{SchLiChe11, *LiMorKle13}, which gives VBM offsets of $1.2\pm0.1\,\eV$ and $0.0\pm0.1\,\eV$ for \PT\ and \ST\ relative to \BT, respectively (also compare \tab{tab:results}).

\Fig{fig:sth}(b,c) shows that STH formation leads to the emergence of a localized atom-like level above the VBM. In the case of \PT\ this O level is noticeably closer to the VBM (see \tab{tab:results}). This suggests that STH formation is energetically less favorable and provides a rationale for the positive STH formation energy in \PT.

It still remains to resolve the origin of the higher lying VBM in \PT\ compared to \BT\ and \ST. To this end, we show in \fig{fig:banddos}(b-d) the band structures of the three materials in question. While the valence band edges of \ST\ and \BT\ are similar in the case of \PT\ one observes a distinct band splitting around the $X$ point, which forms the VBM and gives rise to a tail in the density of states, compare \fig{fig:banddos}(e). The feature is caused by the interaction of Pb $6s$ states, which are spatially rather extended and energetically located just beneath the topmost valence band [see \fig{fig:banddos}(c)], with O $2p$ states, which constitute the top of the valence band. Strong $s-p$ coupling and large upward VBM shifts are also observed for example in Bi compounds \cite{SchLiChe11, *LiMorKle13} for similar reasons.

\begin{figure}
\includegraphics[scale=\myscale]{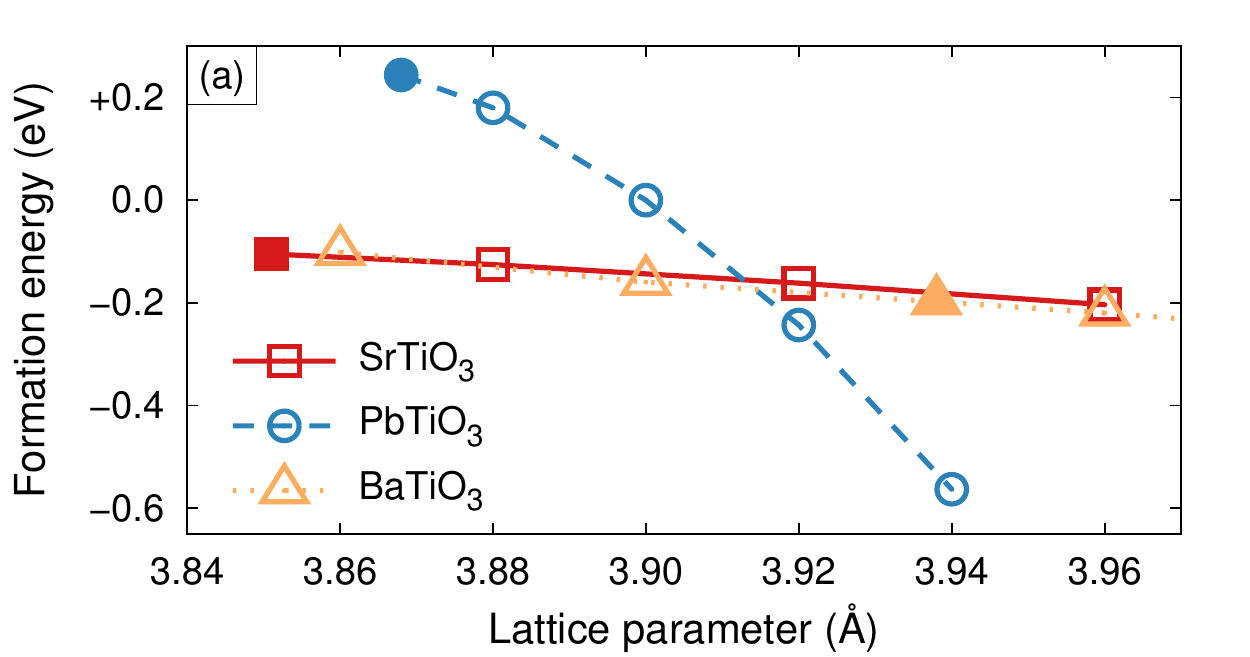}
\includegraphics[scale=\myscale]{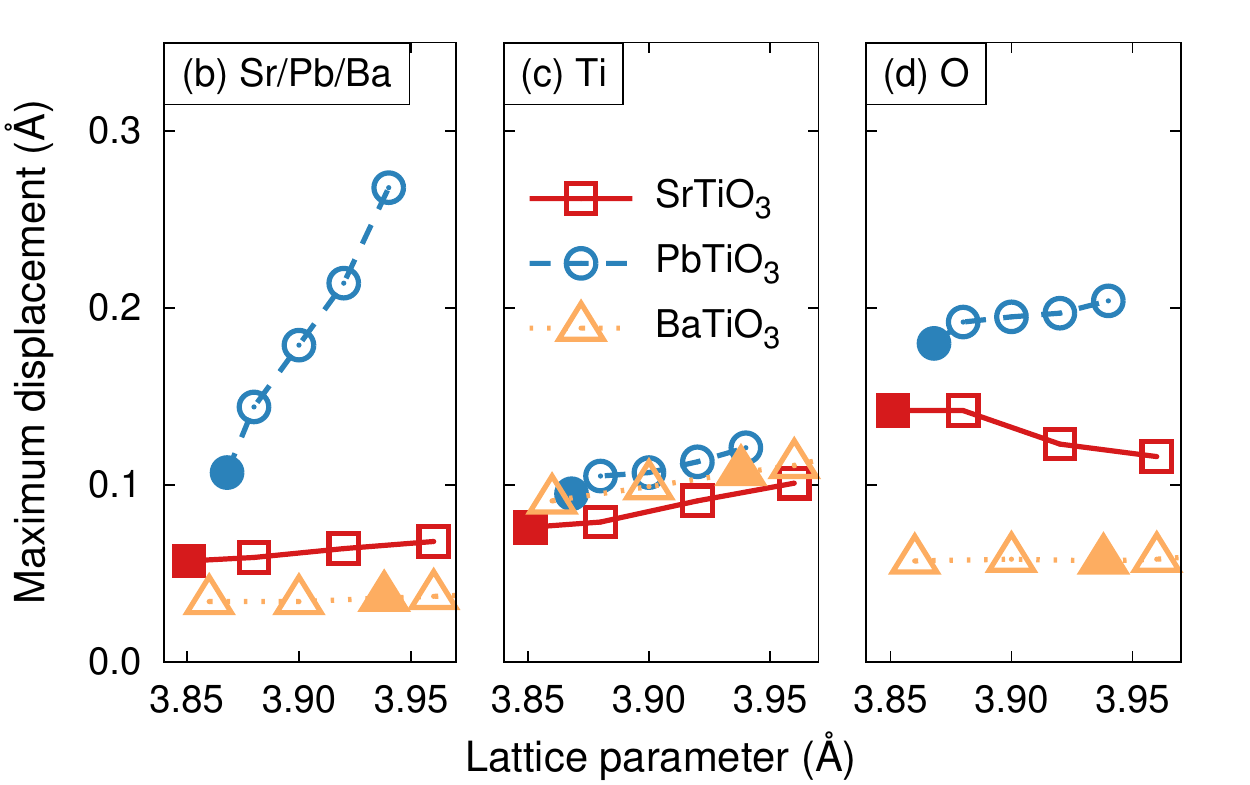}
  \caption{
    (a) STH formation energies and (b-d) maximum atomic displacements per atom type for cubic \ST, \PT, and \BT\ as a function of lattice parameter. Filled symbols indicate data corresponding to the equilibrium lattice constant.
  }
  \label{fig:eform_PT}
\end{figure}

Based on this argument on can thus expect that within this group of materials VBM alignment can serve as a good predictor for STH formation (or absence thereof) in so far as it is reflective of the coupling of low lying cation states to the O $2p$ states comprising the valence band. One can furthermore anticipate that more increasing the lattice constant should lead to stronger self-trapping and thus more negative formation energies since it allows for more ionic relaxation. In the case of \PT\ increasing the lattice constant should additionally reduce $s-p$ coupling resulting in a lower VBM and thus a further reduction of the STH formation energy. This is indeed observed as shown in \fig{fig:eform_PT}(a), which shows the STH formation energies for the three compounds considered in this study as a function of lattice parameter. For \BT\ and \ST\ one obtains similar formation energies and a modest decrease of $\Delta E_f$ with lattice expansion. By contrast \PT\ exhibits a much stronger variation of the formation energy with volume, which is also reflected in the maximum atomic displacements that are show in \fig{fig:eform_PT}(b-c). Particularly apparent is the strong involvement of Pb ions in the STH structure already at the equilibrium lattice parameter, which is further enhanced upon volume expansion.

\section{Conclusions}

In summary we have demonstrated from a methodological standpoint that
({\em i}) the application of LDA+$U$ potentials to O $2p$ states can be physically motivated in the case of self-trapping in oxides,
({\em ii}) suitable parametrizations match the forces and energetics of hybrid functionals well,
and that
({\em iii}) both LDA+$U$ parametrizations and the HSE06 functional are reasonably close to piece-wise linearity in the case of STHs in the oxidic perovskites considered in this study.
In terms of materials relevant results it was shown that hole self-trapping should occur in \ST\ and \BT\ but not \PT. The absence of bulk STHs in the latter case can be explained by the higher VBM compared to the other two materials, which originates from strong coupling between Pb $6s$ and O $2p$ states.
The results described in this paper provide a basis for similar studies in other oxide materials. They illustrate what kind of effects should be taken into account and indicate which fundamental parameters can serve as approximate predictors for the presence or absence of self trapping.

\section*{Acknowledgments}

P.E. acknowledges funding from the {\em Area of Advance -- Materials Science} at Chalmers, the Swedish Research Council in the form of a Young Researcher grant, and the European Research Council via a Marie Curie Career Integration Grant. A. K. acknowledges support by the German Science Foundation via the collaborative research center on electrical fatigue of functional materials (SFB 595). D.{\AA}. and B.S. acknowledge funding from the NA-22 agency. Com\-puter time allocations by the Swedish National Infrastructure for Computing at NSC (Link\"oping) and C3SE (Gothenburg) are gratefully acknowledged.

\end{document}